\documentclass[11pt,tightenlines,nofootinbib,amsfonts,prd,showpacs,preprintnumbers]{revtex4}
\usepackage{graphicx}

\begin{document}

\newcommand {\intsum}{\ \int\!\!\!\!\!\!\!\!\!\sum}
\newcommand {\real}{\mathrm{Re}}
\newcommand {\imag}{\mathrm{Im}}
\newcommand {\Ds} {D\!\!\!\!/}
\newcommand {\cS}{\mathcal{S}}
\newcommand {\cM}{\mathcal{M}}
\newcommand {\cN}{\mathcal{N}}
\newcommand {\cA}{\mathcal{A}}
\newcommand {\cB}{\mathcal{B}}
\newcommand {\cC}{\mathcal{C}}
\newcommand {\cD}{\mathcal{D}}
\newcommand {\cE}{\mathcal{E}}
\newcommand{\Res}[1]{\!\!\!\!\begin{array}{c}\raisebox{-8.5pt}{Res}
  \\{\scriptstyle \ #1}\end{array}\!\!\!\!}
\newcommand {\Tr}{\mathrm{Tr}}
\newcommand {\gf}{g_{\mathrm eff}}

\preprint{TUW-04-34}

\title{Resummed one-loop gluonic contributions to the color superconducting color charge density vanish}
\author{Andreas Gerhold\footnote{Electronic address: gerhold@hep.itp.tuwien.ac.at}}
\affiliation{Institut f\"ur Theoretische Physik, Technische Universit\"at Wien, 
  Wiedner Hauptstr. 8-10, A-1040~Vienna, Austria}

\begin{abstract}
  It is shown that gluonic corrections to the tadpole diagrams  vanish
  in the 2SC and  CFL phases at the order where one might have
  expected NLO corrections. 
  This implies that the gluonic part of the color charge density 
  is negligible at the order of our computation.
  This statement remains true after inclusion of the gluon vertex
  correction and contributions from Nambu-Goldstone bosons. 
\end{abstract}

\maketitle

\section{Introduction}

It is well known that cold dense quark matter is a color superconductor \cite{Rajagopal:2000wf}.
At asymptotic densities, when the QCD coupling constant becomes small because of asymptotic freedom \cite{Gross:1973id}, calculations
from first principles, using the QCD Lagrangian, should be possible. In BCS theory the superconductivity gap 
is $\phi_{BCS}\sim\mu\exp(-c/G^2)$, but it turns out that this result is modified in a color superconductor because
of dynamical screening. Using renormalization group techniques, Son \cite{Son:1998uk} showed
that at leading order the color superconductivity gap is given by $\phi_0\simeq b_0g^{-5}\mu\exp(-3\pi^2/(\sqrt{2}g))$.

\begin{figure}
  \includegraphics{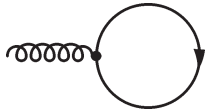}
  \vspace{-2mm}
  {\caption{Leading order tadpole diagram \label{fig0}}}
\end{figure}

In nature color superconducting
phases might exist in the interior of compact stars, if the density is sufficiently high.
Depending on the chemical potential and the strange quark mass, different
phases have been predicted, e.g. CFL \cite{Alford:1998mk}, 2SC \cite{Alford:1997zt}, CFL-K \cite{Schafer:2000ew}, 
g2SC \cite{Shovkovy:2003uu}, gCFL \cite{Alford:2003fq}, 
LOFF \cite{Alford:2000ze}. 
Bulk matter in neutron stars has to be neutral with respect to color (and electric) charges. 
This neutrality condition plays an important role in the determination of the phase structure of
cold dense quark matter.
Most calculations at intermediate densities are based on NJL models, where color neutrality has to be imposed as an 
external condition \cite{Shovkovy:2003uu,Alford:2003fq,Amore:2001uf,Alford:2002kj,Steiner:2002gx}. In QCD, however,
color neutrality is enforced automatically by the dynamics of the gluons \cite{Gerhold:2003js,Kryjevski:2003cu,
Dietrich:2003nu}. The gluon field acquires a 
non-vanishing expectation value, which acts as an effective chemical potential for the color
charge\footnote{
It has been speculated that a non-vanishing expectation value of the gluon field might also be a possibility to
cure the chromomagnetic instability of the gapless color superconducting phases \cite{Huang:2004bg,Casalbuoni:2004tb}.}.
This expectation value can be computed from the Yang-Mills equation \cite{Dietrich:2003nu},
\begin{equation}
  {\delta\Gamma\over\delta A_0^a}\bigg|_{A=\bar A}=0,
\end{equation}
where $\Gamma$ is the effective action, and $\bar A$ is the expectation value of the gluon field.
On the other hand, $\bar A$ can be computed perturbatively from the tadpole diagram \cite{Gerhold:2003js}
\begin{equation} 
  \mathcal {T}_a={\delta\Gamma\over\delta A_0^a}\bigg|_{A=0}
\end{equation}  
by attaching an external gluon propagator.
In the 2SC phase one finds \cite{Gerhold:2003js}
\begin{equation}
   \mathcal {T}_a\simeq-\delta_{a8}{2\over\sqrt{3}\pi^2}g\mu\phi_0^2\log\left({2\mu\over\phi_0^2}\right) \label{tad}
\end{equation}
and \cite{Dietrich:2003nu}
\begin{equation}
  \bar A^0_a\simeq-\delta_{a8}\sqrt{6}\pi^2{\phi_0^2\over g^2\mu}. \label{A0}
\end{equation}
This result comes from the leading order diagram (Fig.~(\ref{fig0})),
which corresponds to the quark part of the color charge density,
\begin{equation}
  \rho^a_{(q)}=\sum_{f=1}^{N_f}\bar\psi_f T^a\gamma_0\psi_f.   \label{rf}
\end{equation}
One might ask whether the gluons could also contribute to the charge density. In 
Ref. \cite{Alford:2002kj} it was argued that the gluonic contribution should vanish, since the gluonic part of the
charge density
\begin{equation}
  \rho^a_{(gl)}=f^{abc}A^b_i F^c_{i0}
\end{equation}
contains the chromoelectric field strength, which vanishes in any (super-)con\-ducting system. The aim
of section \ref{sec1} of this paper is to corroborate this argument by analyzing the tadpole
diagram with a gluon loop (Fig. (\ref{fig1})). We will show that this diagram vanishes (within our approximations)
in the 2SC phase.
In Secs. \ref{sec2} and \ref{sec3} we will show that this diagram still
vanishes if one includes the gluon  vertex correction and contributions from 
Nambu-Goldstone bosons. In Sec. \ref{sec4} we will briefly discuss the CFL phase, which shows the
same qualitative features in this respect. Sec. \ref{sec5} contains our conclusions.


\section{Gluon self energy and gluon loop tadpole (2SC)\label{sec1}}

\begin{figure}
  \includegraphics{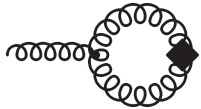}
  \vspace{-2mm}
  {\caption{Tadpole diagram with resummed gluon propagator \label{fig1}}}
\end{figure}

The Lagrangian of QCD is given by
\begin{eqnarray}
  \mathcal{L}_{QCD}&=&\mathcal{L}_1+\mathcal{L}_2,\\
  \mathcal{L}_1&=&\sum_{f=1}^{N_f}\bar\psi_f\left(i\partial\!\!\!/+\mu\gamma_0+gA\!\!\!/^a T^a\right)\psi_f,\\
  \mathcal{L}_2&=&-{1\over4}F_{\mu\nu}^a F^{\mu\nu a}+{1\over2\xi}(\partial^\mu A_\mu^a)^2-\bar c^a\partial^\mu D_\mu^{ab}c^b   .\label{x9}
\end{eqnarray}
For definiteness we have chosen a covariant gauge. The gluon self energy in the 2SC phase can be computed with the
usual Nambu-Gor'kov ansatz for the quark propagator \cite{Rischke:2000qz}. At  high density and zero temperature
only the diagram with a  quark loop contributes to the gluon self energy at leading order. Therefore the gluon self energy
is gauge independent at this order.

Let us consider the
tadpole diagram in Fig. (\ref{fig1}), where the rhombus indicates a
resummed gluon propagator.
In the 2SC phase the gluon self energy has
off-diagonal  components with respect to the color indices, therefore the diagram does not vanish a priori.
The non-vanishing components of the gluon self energy in the 2SC phase are
\begin{equation}
  \Pi_{11}=\Pi_{22}=\Pi_{33}, \quad \Pi_{44}=\Pi_{55}=\Pi_{66}=\Pi_{77}, \quad
  \Pi_{45}=-\Pi_{54}=\Pi_{67}=-\Pi_{76}.
\end{equation}
In Ref. \cite{Rischke:2000qz} the gluon self energy is diagonalized by a unitary
transformation in color space. For our purposes it is more convenient not to perform this
transformation, since we want to keep the structure constants totally antisymmetric.
For the $44$ and $45$ components of the self energy analytical results can be obtained rather easily, as we
will demonstrate in the following. We shall always assume that the energy and the momentum of the gluon are
much smaller than $\mu$, because if there is any contribution to the tadpole diagrams, it will come from soft
gluons\footnote{For $p_0\gg\phi$ the gluon self energy is the same as in the normal phase. Since
it is then proportional to $\delta_{ab}$, it gives no contribution to the tadpole diagram. Furthermore the 
typical gluon momenta relevant for the tadpole diagram would be of the order of the Debye or Meissner masses
(or even smaller).}.
In the notation of Ref. \cite{Rischke:2000qz} we have
\begin{eqnarray}
  &&\!\!\!\!\!\!\!\!\!\!\!\!\!\!\!\!\!\!\!
  \Pi_{44}^{00}=-{1\over2}g^2\int {d^3k\over(2\pi)^3}\sum_{e_1,e_2=\pm}
  (1+e_1e_2\,{\bf\hat k}_1\cdot{\bf\hat k}_2)\nonumber\\
  &&\times\left(n_1^0(1-n_2)+(1-n_1^0)n_2\right) \left[{1\over p_0+\epsilon_1^0+\epsilon_2+i\varepsilon}
  -{1\over p_0-\epsilon_1^0-\epsilon_2+i\varepsilon}
  \right], \label{pi44}
\end{eqnarray}
and
\begin{eqnarray}
  &&\!\!\!\!\!\!\!\!\!\!\!\!\!\!\!\!\!\!\!
  -i\Pi_{45}^{00}\equiv\hat\Pi^{00}=-{1\over2}g^2\int {d^3k\over(2\pi)^3}\sum_{e_1,e_2=\pm}
  (1+e_1e_2\,{\bf\hat k}_1\cdot{\bf\hat k}_2)\nonumber\\
  &&\times \left(n_1^0(1-n_2)-(1-n_1^0)n_2\right)
  \left[{1\over p_0+\epsilon_1^0+\epsilon_2+i\varepsilon}
  -{1\over p_0-\epsilon_1^0-\epsilon_2+i\varepsilon}
  \right]. \label{pi45}
\end{eqnarray}
Here we have used the notations of Ref. \cite{Rischke:2000qz}, 
\begin{equation}
  \epsilon_i=\sqrt{(\mu-e_i k_i)^2+|\phi_i|^2},
\end{equation}
\begin{equation}
  n_i={\epsilon_i+\mu-e_ik_i\over 2\epsilon_i},
\end{equation}
and the superscript ``0'' means $\phi\to0$.
First let us evaluate $\imag \Pi_{44}^{00}$. Without loss of 
generality we assume $p_0>0$. We observe that  $\imag \Pi_{44}^{00}$ vanishes for $p_0<\phi$ because of
$\epsilon_1^0>0$ and $\epsilon_2>\phi$.
For $p_0<\mu$ the only contribution arises from $e_1=e_2=+$. With
${\bf k}_1={\bf k}+{\bf p}$, ${\bf k}_2={\bf k}$, ${\bf\hat k}\cdot{\bf\hat p}=t$, $k_1=\sqrt{k^2+p^2+2pt}$,
$\xi=k-\mu$
we find
\begin{eqnarray}
  &&\!\!\!\!\!\!\!\!\imag\Pi_{44}^{00}=-\Theta(p_0-\phi){g^2\over8\pi}\int_{-\mu}^\infty d\xi\,(\mu+\xi)^2
  \int_{-1}^1dt\left(1+{\mu+\xi+pt\over k_1}\right)\nonumber\\
  &&\quad\times\Theta\!\left(p_0-\sqrt{\xi^2+\phi^2}\right)
  \bigg[{\sqrt{\xi^2+\phi^2}+\xi\over2\sqrt{\xi^2+\phi^2}}\,
  \delta\left(p_0-\mu+k_1-\sqrt{\xi^2+\phi^2}\right)\nonumber\\
  &&\qquad+{\sqrt{\xi^2+\phi^2}-\xi\over2\sqrt{\xi^2+\phi^2}}\,
  \delta\left(p_0+\mu-k_1-\sqrt{\xi^2+\phi^2}\right)\bigg].
\end{eqnarray}
Because of the step function in the second line 
we have $\xi\ll\mu$ for $p_0\ll\mu$. Therefore it is sufficient to
approximate $\phi$ with its value at the Fermi surface, which we denote with $\phi_0$. 
After performing the angular integration we obtain
\begin{eqnarray}
  &&\!\!\!\!\!\!\!\!\imag\Pi_{44}^{00}=-\Theta(p_0-\phi_0){g^2\over16\pi p} 
  \int_{-\sqrt{p_0^2-\phi_0^2}}^{\sqrt{p_0^2-\phi_0^2}} d\xi\,\Theta\!\left(p+p_0-\xi-\kappa\right)
  \Theta\!\left(p-p_0+\xi+\kappa\right)\nonumber\\
  &&\times{1\over\kappa}\left[(4\mu^2-p^2+(p_0+2\xi)^2)(-\xi+\kappa)-\phi_0^2(2p_0+3\xi-\kappa)\right],
  \label{gls9}
\end{eqnarray} 
with $\kappa=\sqrt{\xi^2+\phi_0^2}$.
The $\xi$ integration is now straightforward, and we find
\begin{eqnarray}
  &&\!\!\!\!\!\!\!\!\!\!\!\!\!\!\!\!
  \imag\Pi_{44}^{00}=-\Theta(p_0-\phi_0)\Theta(-p_0+\sqrt{p^2+\phi_0^2})
  {g^2\over24\pi p}\sqrt{p_0^2-\phi_0^2}\left(12\mu^2-3p^2+p_0^2-\phi_0^2\right) \nonumber\\
  &&\quad-\Theta(p_0-\sqrt{p^2+\phi_0^2}){g^2\mu^2\phi_0^2\over 2\pi(p_0^2-p^2)}\nonumber\\
  &&\qquad\times\left[1+{-6\phi_0^2p_0^2(p_0^2-p^2)
  +3(p_0^2-p^2)^3+\phi_0^4(p^2+3p_0^2)\over 12\mu^2(p_0^2-p^2)^2}\right].
\end{eqnarray}
This result is a generalization of the one-loop gluon self energy at zero temperature in the
normal phase given in \cite{Ipp:2003qt}. As in the normal phase, the leading part of the
gluon self energy is of the order $g^2\mu^2$ (for $p_0,p\ll\mu$).
In order to maintain consistency in the following, we shall keep only these leading terms,
since contributions from lower powers of $\mu$ might mix with contributions from diagrams
with a higher number of loops. Denoting this approximation with a tilde, we have
\begin{equation}
  \imag\tilde\Pi_{44}^{00}=-\Theta(p_0-\phi_0){g^2\mu^2\over2\pi}\bigg[\Theta(-p_0+\sqrt{p^2+\phi_0^2})
  {\sqrt{p_0^2-\phi_0^2}\over p}
  +\Theta(p_0-\sqrt{p^2+\phi_0^2}){\phi_0^2\over p_0^2-p^2}\bigg].
\end{equation}
In the following we will refer to this  analogon of the HDL approximation as the ``leading order'' approximation.
$\imag\Pi^{00}_{44}$ is an odd function of $p_0$, therefore the real part can be
calculated with the following dispersion relation\footnote{In principle it is not sufficient to take
only the part of Eq. (\ref{pi44}) where $e_1=e_2=+$ when computing the real part of the gluon self energy.  However,
as argued in \cite{Rischke:2001py},  for $p_0,p\ll\mu$ antiparticles (being always far from the ``Fermi surface'')
only give a constant term in the transverse gluon self energy, and this constant is the same in the
superconducting phase and in the normal phase (at leading order).} \cite{Rischke:2001py}
\begin{equation}
  \real\tilde\Pi_{44}^{00}(p_0,p)={1\over\pi}{\mathcal P}\int_0^\infty d\omega\,\imag\tilde\Pi_{44}^{00}(\omega,p)
  \left({1\over \omega+p_0}+{1\over\omega-p_0}\right).
\end{equation}
In contrast to the $11$, $22$, $33$ and $88$ components of the gluon self energy,
the principal value integral can be performed analytically in this case, 
\begin{eqnarray}
   &&\!\!\!\!\!\!\!\!\!\!\!\!\!\!\!\!\!\!\!\!
  \real\tilde\Pi_{44}^{00}= -{g^2\mu^2\over\pi^2}\Bigg[\Theta\left(\phi_0^2-p_0^2\right)
  \left(1-{\sqrt{\phi_0^2-p_0^2}\over p}
  \arctan{p\over\sqrt{\phi_0^2-p_0^2}}\right)\nonumber\\
  &&+\Theta\left(p_0^2-\phi_0^2\right)\left(1+{\sqrt{p_0^2-\phi_0^2}\over2 p}\log\bigg|{\sqrt{p_0^2-\phi_0^2}-p
  \over\sqrt{p_0^2-\phi_0^2}+p}\bigg|\right)
  +{\phi_0^2\over 2(p^2-p_0^2)}\log\bigg|{\phi_0^2+p^2-p_0^2\over\phi_0^2}\bigg|\Bigg].
\end{eqnarray}
As a consistency check, we may extract the Debye mass in the normal and in the superconducting phase,
\begin{eqnarray}
  -\lim_{p\to0}\lim_{p_0\to0}\lim_{\phi_0\to0}\real\tilde\Pi_{44}^{00}(p_0,p)
  \!\!\!&=&\!\!\!{g^2\mu^2 \over\pi^2},\\
  -\lim_{p\to0}\lim_{p_0\to0}\real\tilde\Pi_{44}^{00}(p_0,p)
  \!\!\!&=&\!\!\!{g^2\mu^2 \over2\pi^2},
\end{eqnarray}
which are the standard results for $N_f=2$ \cite{Rischke:2000qz}. The other Lorentz components of 
$\tilde\Pi_{44}^{\mu\nu}$ may be evaluated in a similar manner \cite{diss}.


The off-diagonal components of the gluon self energy can be evaluated in an analogous way. In place of Eq. (\ref{gls9})
we find
\begin{eqnarray} 
  &&\!\!\!\!\!\!\!\!\imag\hat\Pi^{00}=-\Theta(p_0-\phi_0){g^2\over4\pi p} 
  \int_{-\sqrt{p_0^2-\phi_0^2}}^{\sqrt{p_0^2-\phi_0^2}} d\xi\,\Theta\!\left(p+p_0-\xi-\kappa\right)
  \Theta\!\left(p-p_0+\xi+\kappa\right)\nonumber\\
  &&\times
  {\mu\over\kappa}\left(\phi_0^2-(p_0+2\xi)(\kappa-\xi)\right),
\end{eqnarray}
which gives
\begin{equation}
  \imag\hat\Pi^{00}=-\Theta(p_0-\sqrt{p^2+\phi_0^2})g^2\mu {p_0\phi_0^2(\phi_0^2+p^2-p_0^2)
  \over 2\pi(p_0^2-p^2)^2}.
\end{equation}
We observe that in this one-loop result 
there is a term which is linear in $\mu$, but there is no term of order
$g^2\mu^2$. Therefore we have at our order of accuracy
\begin{equation}
  \imag\tilde{\hat\Pi}^{\raisebox{-5pt}{\scriptsize{00}}}=0.
\end{equation}
In fact this can be seen rather directly from the ($e_1=e_2=+$)-component of Eq. (\ref{pi45}): 
if we set $d^3k\to 2\pi\mu^2d\xi dt$ 
we find at the order $g^2\mu^2$ that the integrand is odd with respect to $\xi\to-\xi$, $t\to-t$, and therefore
the integral vanishes at this order. This argument also holds for the real part,  
$\real\tilde{\hat\Pi}^{\raisebox{-5pt}{\scriptsize{00}}}=0$.
In the same way it can be shown that $\tilde{\hat\Pi}^{\raisebox{-5pt}{\scriptsize{0i}}}=0$
and $\tilde{\hat\Pi}^{\raisebox{-5pt}{\scriptsize{ij}}}=0$.

This implies that the resummed gluon propagator is diagonal in the color indices at leading order. Since the
three-gluon vertex is antisymmetric in the color indices at tree level, 
we conclude that the tadpole diagram of Fig. (\ref{fig1}) vanishes at the order of our computation. This means that
at this order the expectation value of the gluon field, Eq. (\ref{A0}), is
not changed by the tadpole diagram in Fig. (\ref{fig1}).   
Therefore the effective chemical potential is determined completely by the
quark part of the color charge density at the order of our computation, whereas the gluonic part is negligible at this order.


We note that there is no contribution from the tadpole diagram with
a ghost loop. The reason is that there is no direct coupling between quarks and ghosts in $\mathcal{L}_{QCD}$, therefore the
ghost self energy vanishes at the order $g^2\mu^2$, and the ghost propagator is proportional to the unit matrix in color space at leading order.
Furthermore we remark that in
particular the gauge dependent part of the gluon propagator is diagonal in the color indices at leading order.
Therefore the effective chemical potential is gauge independent at this order.

\section{Gluon vertex correction (2SC)\label{sec2}}

\begin{figure}
  \includegraphics{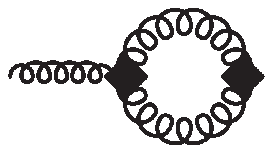} \qquad\quad
  \includegraphics{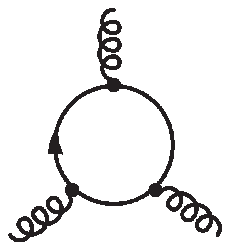}
  \vspace{-1mm}
  {\caption{Tadpole diagram with resummed gluon propagator and three-gluon vertex correction \label{fig2}}}
\end{figure}

We would like to check whether the above result is modified if one replaces the
tree level gluon vertex with the one-loop vertex correction, as shown in Fig. (\ref{fig2}). 
We remark that the tadpole diagram in  Fig. (\ref{fig2}) corresponds to a higher order correction to
the \emph{fermionic} part of the charge density (Eq. (\ref{rf})).
Let us evaluate the one-loop three-gluon vertex of Fig. (\ref{fig2}) in the limit where one of the gluon 
momenta 
approaches zero. 
One finds for instance\footnote{The vertex correction is gauge independent at leading order, since only the diagram with
a fermion loop (Fig.~\ref{fig2}) contributes at leading order.}
\begin{eqnarray}
  &&\!\!\!\!\!\!\!\!\!\!\!
  \Gamma^{000}_{485}(P,0,-P)={ig^3\over2\sqrt{3}}\int {d^3k\over(2\pi)^3}\sum_{e_1,e_2=\pm}
  (1+e_1e_2\,{\bf\hat k}_1\cdot{\bf\hat k}_2)\nonumber\\ 
   &&\times\bigg[\bigg({1\over\left(p_0-\epsilon_1-\epsilon_2^0+i\varepsilon\right)^2}
   +{1\over\left(p_0+\epsilon_1+\epsilon_2^0+i\varepsilon\right)^2}\bigg)
  {2\phi_1^2+3\xi_1^2-3\xi_1\epsilon_1\mathrm{sgn}(\xi_2)
  \over\epsilon_1^2}\nonumber\\
  &&\ \ +\bigg({1\over p_0-\epsilon_1-\epsilon_2^0+i\varepsilon}
   -{1\over p_0+\epsilon_1+\epsilon_2^0+i\varepsilon}\bigg){\phi_1^2\over\epsilon_1^3}\bigg], \label{gls19}\\
  &&\!\!\!\!\!\!\!\!\!\!\!
  \Gamma^{000}_{484}(P,0,-P)= -{g^3\over2\sqrt{3}}\int {d^3k\over(2\pi)^3}\sum_{e_1,e_2=\pm}
  (1+e_1e_2\,{\bf\hat k}_1\cdot{\bf\hat k}_2)\nonumber\\
  &&\times\bigg[\bigg({1\over\left(p_0-\epsilon_1-\epsilon_2^0+i\varepsilon\right)^2}
   +{1\over\left(p_0+\epsilon_1+\epsilon_2^0+i\varepsilon\right)^2}\bigg)
  {-\mathrm{sgn}(\xi_2)(2\phi_1^2+3\xi_1^2)+3\xi_1\epsilon_1\over\epsilon_1^2}\nonumber\\
  &&\ \ -\bigg({1\over p_0-\epsilon_1-\epsilon_2^0+i\varepsilon}
   -{1\over p_0+\epsilon_1+\epsilon_2^0+i\varepsilon}\bigg){\mathrm{sgn}(\xi_2)\phi_1^2\over\epsilon_1^3}\bigg],
  \label{gls21}
\end{eqnarray}
where ${\bf k}_1={\bf k}+{\bf p}$, ${\bf k}_2={\bf k}$ and $\xi_i=e_ik_i-\mu$.
The integrals can be evaluated in a similar way as in the previous section. 
It is easy to see that $\Gamma^{\mu0\nu}_{484}(P,0,-P)$ vanishes at the order $g^3\mu^2$.
In general \cite{diss,Casalbuoni:2001ha} one finds that 
$\Gamma^{\mu0\nu}_{abc}(P,0,-P)$ is non-vanishing at the order $g^3\mu^2$ only for those combinations of color
indices where $f_{abc}$ is non-vanishing. In the previous section we have seen that the gluon 
propagator is diagonal in the color indices at leading order.
Therefore the tadpole diagram in Fig. (\ref{fig2}) also vanishes at this order.

\section{Mixing with Nambu-Goldstone bosons (2SC)\label{sec3}}

The symmetry breaking pattern of the 2SC phase is  $SU(3)_c\times U(1)_B\to SU(2)\times\tilde U(1)_B$ 
\cite{Alford:1997zt}.
Therefore five massless Nambu-Goldstone (NG) bosons appear in this phase, which correspond to fluctuations of the
diquark condensate. 
As shown in \cite{Casalbuoni:2000cn,Miransky:2001sw,Rischke:2002rz}, one finds the following 
effective action for gluons,ghosts and NG bosons after integrating out the quarks, 
\begin{eqnarray}
  &&\!\!\!\!\!\!\!\!
  \Gamma=\int d^4x\mathcal{L}_2+{1\over2}\Tr\log\left({\mathcal S}^{-1}+\gamma_\mu\Omega^\mu\right)
  \nonumber\\
  &&=\int d^4x\mathcal{L}_2
  +{1\over2}\Tr\log{\mathcal S}^{-1}+{1\over2}\Tr\left({\mathcal S}\gamma_\mu\Omega^\mu\right)
  -{1\over4}\Tr\left({\mathcal S}\gamma_\mu\Omega^\mu{\mathcal S}\gamma_\nu\Omega^\nu\right)\nonumber\\
  &&\quad+{1\over6}\Tr\left({\mathcal S}\gamma_\mu\Omega^\mu{\mathcal S}\gamma_\nu\Omega^\nu
  {\mathcal S}\gamma_\rho\Omega^\rho\right)+\ldots, {\label{seff}}
\end{eqnarray}
where $\mathcal{L}_2$ is given in Eq. (\ref{x9}), the quark propagator reads, in the notation of Ref.
\cite{Rischke:2002rz},
\begin{equation}
   \mathcal S=\left(
    \begin{array}{cc} [G_0^+]^{-1}
    & \Phi^- \\
    \Phi^+ & [G_0^-]^{-1}
    \end{array}
  \right)^{-1}, 
\end{equation}
and $\Omega_\mu$ is given by
\begin{equation}
   \Omega^\mu(x,y)=-i\left(
    \begin{array}{cc} \omega^\mu(x)
    & 0 \\
    0 & -(\omega^\mu)^T(x)
    \end{array}
  \right)\delta^4(x-y).
\end{equation}
Here $\omega^\mu$ is the (Lie algebra valued) Maurer-Cartan one-form introduced in \cite{Casalbuoni:2000cn},
\begin{equation}
  \omega^\mu={\mathcal V}^\dag\left(i\partial^\mu+g A_a^\mu T_a\right){\mathcal V},
\end{equation}
where
\begin{equation}
  {\mathcal V}=\exp\left[i\left(\sum_{a=4}^7\varphi_aT_a+{1\over\sqrt{3}}\varphi_8 B\right)\right]
\end{equation}
parametrizes the coset space $SU(3)_c\times U(1)_B/SU(2)\times\tilde U(1)_B$ \cite{Miransky:2001sw}, 
$B=({\bf 1}+\sqrt{3}T_8)/3$ is a generator orthogonal to the one of $\tilde U(1)_B$, and
$\varphi_a$ are the NG bosons.
We may expand $\omega^\mu$ in powers of the fields,
\begin{eqnarray}
  &&\omega^\mu=-{1\over\sqrt{3}}\partial^\mu\varphi_8B+\bigg[g A^\mu_a-\partial^\mu\tilde\varphi_a
  -gf_{abc}A^\mu_b\tilde\varphi_c-{g\over3}f_{ab8}A^\mu_b\varphi_8
  -{1\over2}f_{abc}\tilde\varphi_b\partial^\mu\tilde\varphi_c-{1\over6}f_{a8c}\varphi_8\partial^\mu\tilde\varphi_c\nonumber\\
  &&\qquad\quad-{1\over6}f_{ab8}\tilde\varphi_b\partial^\mu\varphi_8+\ldots\bigg]T_a
\end{eqnarray}
where $\tilde\varphi_a\equiv\varphi_a$ for $a=4,5,6,7$ and $\tilde\varphi_a\equiv0$ otherwise.

Let us examine the various terms in the effective action (\ref{seff}). First consider the term 
linear in $\Omega$, which contains the  one-loop gluon tadpole\footnote{The $\varphi$-tadpole vanishes because 
the external external momentum of the tadpole diagram is 
zero.} (Fig. (\ref{fig0})).
In principle this term also gives contributions to higher $n$-point functions, since $\Omega$ contains arbitrarily high powers
of $\varphi$. In particular, there is a contribution to the 
bosonic self energy, but this effect is negligible at leading order since the one-loop tadpole (Eq. (\ref{tad}))
is only linear in $\mu$.

The term quadratic in $\Omega$  contains the bosonic self energy. As in section \ref{sec2} one finds that at leading order
the self energy is diagonal in the color indices. There are mixed terms between gluons and NG bosons, which
could be eliminated by choosing a suitable t'Hooft gauge \cite{Rischke:2002rz}. This is not necessary for our purposes,
but we note that choosing this t'Hooft gauge would not alter our conclusion, since also in the t'Hooft gauge of Ref. \cite{Rischke:2002rz}
the gluon, NG and ghost propagators are diagonal in the color indices at leading order. The gluon and NG propagators in the 
t'Hooft gauge are given explicitly in Eqs. (54) and (55) of Ref. \cite{Rischke:2002rz}. At leading order the propagators
(and in particular the gauge dependent parts of the propagators)
are diagonal in the color indices also without the unitary transformation in color space that is employed in Ref. \cite{Rischke:2002rz}.

The term quadratic in $\Omega$ also contains  
new types of three-boson vertices  proportional to $f_{abc}$, which involve at least one NG boson.

The term cubic in $\Omega$ is only non-vanishing at the order $g^3\mu^2$ 
for those color indices where $f_{abc}\neq0$. In this work
we do not consider vertices with more than three legs, since these would appear only in two-loop 
tadpole diagrams.

To summarize, we find at leading order as in the previous Section that the bosonic propagators are
diagonal in the color indices, and that three-boson vertices of the type  $V^{\mu0\nu}_{abc}(P,0,-P)$ are non-zero 
only for those combinations of 
color indices where $f_{abc}\neq0$. We conclude that the tadpole diagram in Fig. (\ref{fig2}) vanishes
at the order of our computation, even if we take into account NG bosons.

\section{Some remarks on the CFL phase \label{sec4}}
In the CFL phase the one-loop tadpole diagram in Fig. (\ref{fig0}) vanishes if the strange quark mass is zero.
For finite $m_s$ one finds for the tadpole diagram in Fig. (\ref{fig0}) \cite{Gerhold:2004ja}
\begin{equation}
  {\mathcal T}_a\simeq -\delta_{a8}{g\mu m_s^2\over18\sqrt{3}\pi^2}(21-8\ln2).
\end{equation}

In the case $m_s=0$ the situation
for the diagrams in Figs. (\ref{fig1}), (\ref{fig2}) is similar as in the 2SC phase.
In the CFL phase the one-loop gluon self energy is proportional to the unit matrix in color space \cite{Rischke:2000ra}
(as in the normal phase). Therefore the
diagram in Fig. (\ref{fig1}) vanishes identically. 
For the gluon vertex correction the color structure is also similar to the normal phase.
A straightforward (but somewhat lengthy) calculation \cite{diss} shows that
the three-gluon vertex correction $\Gamma_{abc}^{\mu0\nu}(P,0,-P)$
contains a term proportional to $f_{abc}$ and one proportional to $d_{abc}$, but only the first one is 
non-vanishing at the order $g^3\mu^2$. 
The structure of effective action for gluons and NG bosons is similar to the 2SC phase  
\cite{ Casalbuoni:1999wu, Son:1999cm, Zarembo:2000pj, Manuel:2000wm}. 
Thus we find that the tadpole diagrams in the CFL phase
vanish at the order of our computation.

\section{Conclusions \label{sec5}}

In this paper we have shown that corrections to the gluon tadpole in the 2SC and in the 
CFL phases vanish at the order where one might have expected NLO corrections. This means that
at the order of our computation the expectation value of the gluon field, Eq. (\ref{A0}), is
not changed by the tadpole diagrams in Figs. (\ref{fig1}) and (\ref{fig2}).   

Several straightforward extensions of the present work are possible.
For instance one could include photons, which we have neglected since $e\ll g$. 
One could also compute the tadpole diagrams at finite temperature, and for other color superconducting
phases. It would be more challenging, but certainly interesting, to go beyond the approximations of
this work, and to compute higher order corrections to the tadpole.
The fact that the gluonic contributions to the charge density are negligible at the order of our computation
could perhaps also be viewed as a justification of NJL model calculations,
where no gluons are present in the color charge density from the outset. 

\acknowledgments
I would like to thank K. Rajagopal for the suggestion to analyze the gluonic contribution to the
color charge density. I would also like to thank A. Rebhan for interesting discussions.
This work has been supported by the Austrian Science Foundation FWF, project no. P16387-N08.

\end{document}